\documentclass{aa}
\usepackage{graphics}
\usepackage{rotating}

\begin{document}

   \thesaurus{08    % A&A Section 8: Stars
              ( 08:16:6;		% pulsars: general
		08:16:7 PSR B1951+32;	% pulsars: individual
		08.14.1;		% Stars: neutron
		09.19.2;		% ISM: supernova remnants
		09.09.1 CTB 80;		% ISM: individual objects
		05.01.1;		% Astrometry
               )}

   \title{Detection of new Optical Counterpart candidates to PSR B1951+32 with HST/WFPC2}

   \subtitle{}

   \author{R. F. Butler
          \inst{1}
          \and
	  A. Golden\inst{1}\fnmsep
	  \and
          A. Shearer\inst{1}\fnmsep
          }

   \offprints{R. F. Butler, ray@physics.nuigalway.ie}

   \institute{National University of Ireland, Galway, University Road,
Galway, Ireland\\ email: ray@physics.nuigalway.ie
	}

   \date{}

   \maketitle

%________________________________________________________________

   \begin{abstract}

There remain several definitive $\gamma$-ray pulsars that are as yet
undetected in the optical regime. A classic case is the pulsar PSR
B1951+32, associated with the complex CTB 80 SNR. Previous ground
based high speed 2-d optical studies have ruled out candidates to
$m_{V}$ $\sim$ 24.  Hester (2000a) has reported an analysis of archival
HST/WFPC2 observations of the CTB 80 complex which suggest a compact
synchrotron nebula coincident with the pulsar's radio position.
Performing a similar analysis, we have identified a possible optical
counterpart within this synchrotron nebula at $m_{V}$ $\sim$ 25.5 - 26, and
another optical counterpart candidate nearby at $m_{V}$ $\sim$24.5.  We
assess the reality of these counterpart candidates in the context of
existing models of pulsar emission.


   \end{abstract}

%________________________________________________________________

\section{Introduction}

The detection of nonthermal high energy magnetospheric emission from
isolated pulsars has remained a non-trivial problem, despite great
advances in instrumentation and technological expertise.  To date,
only 7 optical pulsars have been detected with emission believed to be
magnetospherically dominated, and despite considerable effort, only 8
$\gamma$-ray pulsars.  In contrast to radio emission, which is
generally believed to be generated in close proximity to the magnetic
poles, no clear theoretical model construct exists as regards the
higher energies. The two principal schools of thought place
$\gamma$-ray emission localised either to the magnetic poles
(Daugherty \& Harding 1996) or located further out in the
magnetosphere (Romani 1996). Considerable problems remain with these
two frameworks, in terms of predicted fluxes, spectral indices and
light curve morphologies, and it is clear that further work is
required on this subject.  This is all the more relevant when one
attempts to address the growing empirical database of lower energy
emission, in particular in the optical regime. A consequence of
non-linear processes within the magnetosphere, this synchrotron
emission forms a useful constraint with which one can attempt to
comprehensively develop a self-consistent theoretical
framework. Consequently it is important to try and acquire synchrotron
(optical/X-ray) data of known $\gamma$-ray pulsars, and so extend this
empirical database.

The pulsar PSR B1951+32, located within the complex combination
supernova remnant (SNR) CTB 80, was first identified as a
steep-spectrum, point-like source in the radio (Strom 1987), and
discovery of radio pulses with an unusually fast 39.5-s period quickly
followed (Kulkarni et al. 1988).  Canonically, the pulsar's age and
the estimated dynamical age of the SNR are consistent at $\sim10^5$
yrs (Koo et al. 1990) and both have been determined to be at a
distance of $\sim$ 2.5 kpc.  There is thus general agreement that the
association is valid.  Evidence for pulsed emission was subsequently
found in $\gamma$-rays (Ramanamurthy et al. 1995) and possibly in X-rays
(Safi-Harb et al. 1995; Chang \& Ho 1997), with upper limits in the
infrared (Clifton et al. 1988). The ROSAT observations
in the X-ray regime do indicate a complex light curve strongly
dominated by the intense X-ray radiation of a pulsar-powered
synchrotron nebula (Safi-Harb et al. 1995; Becker \& Truemper
1996). The unambiguous double-peaked $\gamma$-ray (EGRET) light curve
obtained by Ramanamurthy et al.  (1995) at the appropriate spin-down
ephemeris suggested that the pulsar had a conversion efficiency, in
terms of rotational energy to $\gamma$-rays, of $\sim$
0.004. Consequently there are strong grounds for the possibility of an
optical detection.

However, the pulsar is quite distant and located within a rather
complex SNR. Recent radio observations at 92cm (Strom \& Stappers
2000) indicate that the pulsar is located at the edge of the flat
radio spectrum 'core' of the SNR, and it is clear from X-ray
observations that the pulsar is to some extent interacting with its
environment.  From models such as Pacini \& Salvati (1987) and Shearer
\& Golden (2001) we would expect emission in the $V$ magnitude range
24-26 depending upon the line of sight absorption, estimated from
E(B-V) = 0.8-1.4 (Blair et al. 1984). Ground-based CCD observations by
Blair \& Schild (1985) and Fesen \& Gull (1985) yielded a relatively
crowded field with two possible counterparts at $m_v ~\sim$ 20 and
$m_v ~\sim$ 21 respectively (known hereafter as counterparts 1 and 2).
 
\begin{figure} 
\resizebox{\hsize}{!}{\includegraphics{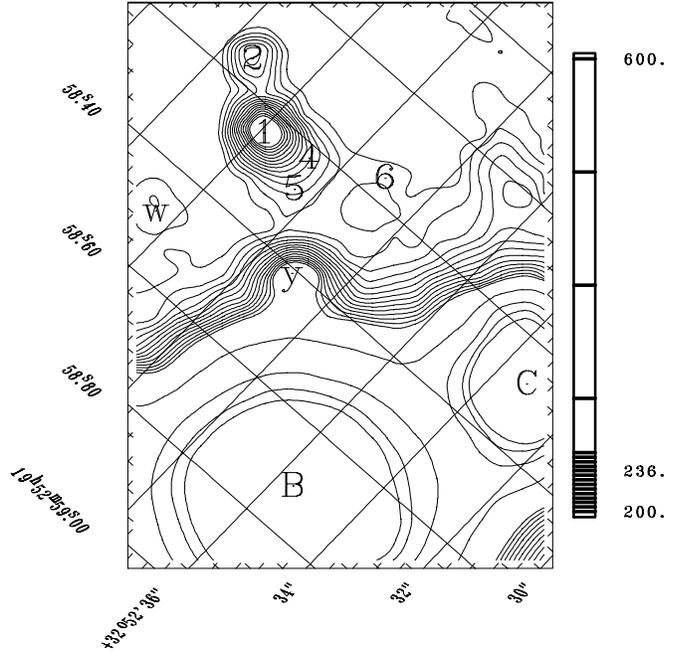}}

\caption{Contour plot of the TRIFFID/MAMA V-band 8173-second summed
image of the central region of CTB 80. This is a re-presentation of
the data first discussed by O'Sullivan et al. (1998); the image has
been rotated and registered to the same coordinate system as the WFPC2
F547M image discussed in this paper. The main contours range from
intensities of 200 to 252 in steps of 3 and were chosen to highlight
the fainter structures. The pulsar optical counterpart candidates from
Blair \& Schild (1985), Fesen \& Gull (1985), and O'Sullivan et
al. (2000) are marked by their ID numbers (1,2,4,5,6). Other nearby
detected stars are also marked (w, y). The bright stars identified as
B and C by Blair \& Schild (1985) are also marked and highlighted by a
few extra contours between intensities of 300 and 600. }

\label{mama_V_ext} 
\end{figure}

Using a ground-based MAMA detector in the TRIFFID camera, we have
previously examined the central field of CTB 80, but could find no
evidence of pulsations in either $B$ or $V$ from these two counterparts
(O'Sullivan et al., 1998).  It was noted that counterpart 1 had an
extension (see Fig.~\ref{mama_V_ext}) towards the mapped radio timing
position given in Table~\ref{psr_radio}, which implied an unresolved
stellar combination and/or plerionic/remnant material. Using
PSF-fitting and deconvolution techniques, the removal of counterpart 1
yielded a best-estimate imposed decomposition of the ``extension''
into two further point sources (numbered 4 and 5 in
Fig.~\ref{mama_V_ext}), while a further point source (numbered 6 in
Fig.~\ref{mama_V_ext}) was detected some distance further away.  These
3 sources had apparent magnitudes of $m_V$ $\approx$ 22.1/$B-V$
$\approx$ 0.8, $m_V$ $\approx$ 22.6/$B-V$ $\approx$ 0.6, and $m_V$
$\approx$ 23.1/$B-V$ $\approx$ 0.0. However, none of them exhibited
pulsed emission at the 1\% level.  As was noted, both the photometry
and time series analysis were complicated by their low signal-to-noise
and proximity to counterpart 1.

It is clear that in order to unambiguously resolve the various apparent
components within the radio error ellipse, diffraction-limited
photometry must be obtained, which would facilitate further attempts
to detect optical pulsations from the suspected optical counterpart(s)
to PSR B1951+32. Consequently we obtained from the HST archive WFPC2
images of the CTB 80 SNR obtained with the F547W, F673N, F656N, and
F502N passbands.  In the next section we detail the reduction of the
various exposures and their astrometric and photometric analyses. From
this we assess the feasibility that we have identified new plausible
optical counterparts, and assess the implications - both in terms of
follow-up high speed 2-d photometry, and for current theoretical
models.

%______________________________________________________________

\section{Analysis of archival HST/WFPC2 observations}

We obtained from the HST archive all existing WFPC2 data of the CTB 80
SNR, as listed in Table~\ref{hst_archive}. The core of the CTB 80
remnant lies on chip WF3 of the WFPC2 camera in every case. Image
processing and photometry were performed using the IRAF (Tody 1993),
STSDAS, and DAOPHOT-II (Stetson 1994) packages. The images in each
band and chip were stacked and cleaned of cosmic rays and hot pixels using
standard techniques. The F547M intermediate-width band enabled us to
perform a deep photometric search for faint stellar sources, to S/N=2
at MAG$_{F547M}$ = 26.7.  Photometry was performed with
DAOPHOT-II/allstar PSF-fitting, applying the correct ``synthetic''
zeropoint \& gain ratio (Holtzman et al. 1995), correcting the
photometry for CTE effects using the recipe of Stetson (1998),
correcting for geometric distortion across the different WF chips, and
correcting to the ``standard'' $0^{\prime\prime}.5$ radius
photometry.

\begin{table}	
\small
\begin{tabular}{cccc}
\hline\noalign{\smallskip}
Date & Filter & Total Exptime  & Notes \\
(dd/mm/yy) & Name & (seconds) & \\
\noalign{\smallskip}
\hline
\noalign{\smallskip}
2/10/97 & F656N & 5300 & H II \\
2/10/97 & F673N & 5400 & S II \\
2/10/97 & F502N & 5400 & O III \\
2/10/97 & F547M & 2600 & Str$\rm\ddot{o}$mgren $y$ \\
\noalign{\smallskip}
\hline
\end{tabular}
\caption{List of WFPC2 observations of the CTB 80 SNR, obtained from the ST-ECF HST archive.}
\label{hst_archive}
\end{table}

\subsection{The Astrometric Solution} 

\label{astrom}

Accurate mapping of the radio positions to the WFPC2 images requires a
more precise astrometric plate solution than is provided in the WFPC2
image headers. The average absolute pointing error of HST is
0$\farcs$8 $\rightarrow$ 1$\farcs$5 (Biretta et al. 2000), due to
errors in the Guide Star Catalog (GSC).  Spacecraft roll can cause an
additional shift of up to 1$\farcs$5. This degree of uncertainty, in
this relatively crowded field, would make it impossible to pin down
the identification of a probable optical counterpart to PSR B1951+32.

\begin{figure}
\resizebox{\hsize}{!}{\includegraphics{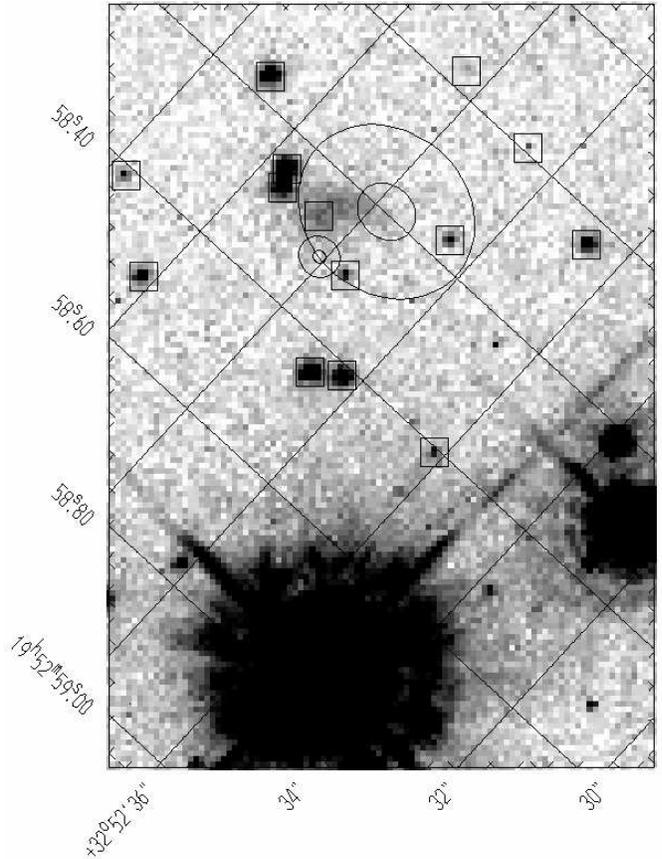}} 

\caption{Section of the WFPC2 F547M image, showing the same field and
orientation as Fig.~\ref{mama_V_ext}.  The coordinate grid shows the
improved astrometric calibration after referencing to the 2MASS
point-source catalog. The mapped PM-corrected radio positions for PSR
B1951+32 from Table~\ref{psr_radio} are marked by black ellipses, the
semi-major and semi-minor axes of which are determined by the
1-$\sigma$ and 3-$\sigma$ total positional uncertainties in RA and Dec
- ie. the quoted error on the radio position, combined with the total
rms error on the HST-2MASS fit for 31 stars. The larger pair of
ellipses marks the interferometric position. The positions derived by
DAOPHOT-II/allstar PSF-fitting for all measured point sources within
5$\farcs$0 of the centre of the synchrotron nebula are indicated by
black squares.}

\label{hst_radio_newIDs}
\end{figure}

Therefore, using 31 stars on chip WF3 in common with the new 2MASS
Point Source Catalog\footnote{2MASS Second Incremental Data Release,
2000 March 2}, and an iterative matching and fitting process, we
derived a much improved astrometric solution with a total rms error of
only 0$\farcs$15.  The various published radio coordinates for PSR
B1951+32 from the literature were filtered to yield only those which
were judged to be fully independent of each other, and based either on
genuine interferometric observations or on a timing solution which had
not been affected by a glitch (although susceptible to timing noise).
We list these two ``best'' radio positions for PSR B1951+32, from
Foster et al. (1990) and Foster et al. (1994), in
Table~\ref{psr_radio}. Given that Migliazzo et al. (2002) have reported
an accurate proper motion for the pulsar of 25 $\pm$ 4 mas yr$^{-1}$
at PA = 252$^{\circ}$ $\pm$ 7$^{\circ}$ from new VLA observations, the
difference between the radio and HST epochs warranted a correction for
significant proper motion. The 10.2 yr difference for the radio timing
position resulted in a correction of $0^{\prime\prime}.254$ in a WSW
direction; the interferometric position was corrected by
$0^{\prime\prime}.22$ in the same direction, due to its 8.7 yr
difference. The resulting positions (see Table~\ref{psr_radio}) were
then mapped onto the improved HST optical astrometric solution, and
their corresponding 1-$\sigma$ and 3-$\sigma$ error ellipses are shown
overplotting the F547M image in Fig.~\ref{hst_radio_newIDs}. As Foster
et al. (1994) note, the discrepancy between them is probably the
result of timing residuals, which may also lead to an underestimate of
the errors for the timing result. This must be borne in mind when
evaluating the nature of objects discovered optically in the vicinity
of these radio positions.

\begin{table*}	% Use table* for 2-col-wide table
\small
\begin{tabular}[b]{lllllllll}
\hline\noalign{\smallskip}
\# & RA & error & Dec & error & Epoch & Equinox & Method & Reference \\   
\  & hh:mm:ss.sss & s.sss & dd:mm:ss.ss & s.ss & & & & \\
\noalign{\smallskip}
\hline
\noalign{\smallskip}
1 & 19:52:58.191 & (0.038) & 32:52:40.22 & (0.56) & 1989.03 & J2000 & Interferometry & Foster et al., 1990  \\
 & 19:52:58.174 & & 32:52:40.15 & & 1997.753 & J2000 & & Foster et al., 1990 - PM corrected \\
2 & 19:52:58.3076 & (0.0051) & 32:52:40.569 & (0.085) & 1987.575 & J2000 & Timing & Foster et al., 1994 \\
 & 19:52:58.2886 & & +32:52:40.49 & & 1997.753 & J2000 & & Foster et al., 1994 - PM corrected \\
\noalign{\smallskip}
\hline
\end{tabular}
\caption{Selected published radio coordinates for PSR B1951+32, which
were used to determine its position on the WFPC2 F547M
image. They are shown both in original form and corrected for proper
motion to the 1997.753 epoch of the HST observations. }
\label{psr_radio}
\end{table*}

\subsection{A Synchrotron Nebulosity Associated with PSR B1951+32} 

In Fig.~\ref{hst_radio_newIDs}, we note the presence of a patch of
nebulosity $\sim$ 0$\farcs$8$\times$1$\farcs$3 in extent, some or all
of which lies within the mapped radio 3-$\sigma$ error ellipses in the
F547M image. It is not immediately evident in the narrowband
images. As noted in section 1, previous ground based time-resolved
MAMA observations (O'Sullivan et al. 1998) have suggested the presence
of such a broadband (present in both $B$ and $V$) extended luminosity
at this location. It was then assumed to be composed of two adjacent
stars (numbered 4 and 5 in Fig.~\ref{mama_V_ext}), but with the
hindsight afforded by these higher resolution HST observations, we now
know that ``star'' 4 marks the compact nebula, while ``star'' 5 is
comprised of a blend of some of this nebula with some of star
4$_{HST}$ in Table~\ref{hst_mags}. Leaving the source intact as a
single compact nebula would have yielded photometry of $m_V$ $\approx$
21.6/$B-V$ $\approx$ 0.7 in that MAMA data. After correction for
reddening of E(B-V) = 1.0 (adopted from O'Sullivan et al. (1998)), the
colour of this nebula is (B-V)$_0$ = -0.3$\pm$0.3; clearly this is a
blue object within the range of these wavebands. This colour is also
consistent with the colours of synchrotron plerions associated with
young pulsars - the Crab and PSR0540-69 in the LMC - with their
approximately flat optical spectral indices.

\subsection{Evidence for an Optical Counterpart to PSR B1951+32} 

Further evidence of the limits of ground-based observations of this
field is seen in Fig.~\ref{hst_radio_newIDs}, where candidate 1 from
Blair \& Schild (1985) is itself resolved into 2 stars, 2$_{HST}$ and
3$_{HST}$, of roughly equal magnitude. Comparison with
Fig.~\ref{mama_V_ext} shows that star 6 from O'Sullivan et
al. (1998) is clearly recovered in this HST data as star 5$_{HST}$, as
are the other nearby faint/crowded stars denoted 'w' and 'y' (the
latter is also resolved into a close pair).

\begin{table}	
\small
% Table FINAL/CTB80_Sy_3.als.3.central.tab  Thu 12:09:59 13-Dec-2001
\begin{tabular}{ccccc}
\hline\noalign{\smallskip}
 \# & RA (2000) & Dec (2000) & Dist. & MAG$_{F547M}$ \\
  & hh:mm:ss.ss & dd:mm:ss.s & arcsec & magnitudes\\
\noalign{\smallskip}
\hline
\noalign{\smallskip}

% ------------------------------------------------------------------------------------------------------

%
\null              1$_{HST}$ &  19:52:58.24 &  32:52:41.0 &         0.20 &        24.26  $\pm$    0.30 \\
\null              2$_{HST}$ &  19:52:58.25 &  32:52:41.8 &         0.97 &        22.49  $\pm$    0.07 \\
\null              3$_{HST}$ &  19:52:58.23 &  32:52:42.0 &         1.11 &        21.69  $\pm$    0.07 \\
\null              4$_{HST}$ &  19:52:58.28 &  32:52:39.9 &         1.16 &        24.54  $\pm$    0.12 \\
\null              5$_{HST}$ &  19:52:58.14 &  32:52:39.0 &         2.23 &        24.41  $\pm$    0.18 \\
\null              6$_{HST}$ &  19:52:58.14 &  32:52:43.4 &         2.69 &        22.18  $\pm$    0.07 \\
\null              7$_{HST}$ &  19:52:58.42 &  32:52:39.2 &         2.90 &        21.96  $\pm$    0.06 \\
\null              8$_{HST}$ &  19:52:58.39 &  32:52:38.7 &         2.95 &        22.30  $\pm$    0.07 \\
\null              9$_{HST}$ &  19:52:58.48 &  32:52:42.6 &         3.55 &        23.33  $\pm$    0.09 \\
\null             10$_{HST}$ &  19:52:57.95 &  32:52:40.8 &         3.57 &        26.06  $\pm$    0.26 \\
\null             11$_{HST}$ &  19:52:58.39 &  32:52:44.1 &         3.73 &        24.59  $\pm$    0.14 \\
\null             12$_{HST}$ &  19:52:57.97 &  32:52:39.1 &         3.79 &        26.20  $\pm$    0.47 \\
\null             13$_{HST}$ &  19:52:58.02 &  32:52:37.1 &         4.68 &        23.24  $\pm$    0.07 \\
\null             14$_{HST}$ &  19:52:58.38 &  32:52:36.6 &         4.74 &        24.77  $\pm$    0.18 \\

% ------------------------------------------------------------------------------------------------------

\noalign{\smallskip}
\hline
\end{tabular}
\caption{The positions and magnitudes derived by DAOPHOT-II/allstar
PSF-fitting for point sources detected in the WFPC2 F547M image of CTB
80. The ``Dist.'' column contains the distance from from each source to
the centre of the synchrotron nebula; only sources within 5$\farcs$0
of this position are listed here.}
\label{hst_mags}
\end{table}

Now we turn our attention to the identification of candidates for an
optical counterpart to PSR B1951+32.  Five point-like sources were
measured within or immediately outside the area covered by the larger
3-$\sigma$ position ellipses. Of these, stars 1$_{HST}$ and 4$_{HST}$
are the most likely to be candidate counterparts. Both lie inside the
3-$\sigma$ interferometric position ellipse, and although both lie
outside the 3-$\sigma$ timing position ellipse, they are closer to it
than any other sources. On the basis of position, neither is clearly a
stronger candidate than the other. The difference between the two
radio coordinates is primarily in Right Ascension - they are
consistent with each other in Declination - which might indicate some
systematic error in RA for one of these radio measurements. This would
then favour the optical counterpart candidate which differed least in
Dec from the two radio positions. However, both candidates lie roughly
equidistant in Declination from a line connecting the two radio
coordinates.  The other possibility, as noted in Section~\ref{astrom},
is that the entire error ellipse may be underestimated for the timing
position; for example, a doubling of the axis lengths of this
3-$\sigma$ ellipse would be just enough to make both candidates
consistent with it.

Only one other measured source lies within the 3-$\sigma$
interferometric position ellipse - star 5$_{HST}$, but this can be
ruled out as a candidate counterpart because it is completely
inconsistent with the radio timing position, even allowing for
possible underestimated errors in the latter.  Finally, the close pair
of stars 2$_{HST}$ and 3$_{HST}$ lie just outside the 3-$\sigma$
interferometric ellipse, and well outside the 3-$\sigma$ timing
position ellipse, which would be sufficient grounds for ruling them
out as candidate counterparts; but two other factors rule them out
with certainty: firstly they are too bright to be consistent with the
emission predicted by the phenomenological models, and secondly they
failed to exhibit optical pulsations in the study by O'Sullivan et
al. (1998).

Of the two candidate counterparts, object 4$_{HST}$ is a
straightforward point-source measurement. However, object 1$_{HST}$
requires some further analysis and comment.  In {\it automatically}
determining and fitting the list of stars detectable in the WFPC2
F547M image, this point-like source was detected within the small
nebulous patch and measured at MAG$_{F547M}$ = 24.3 $\pm$ 0.3
(ie. with a $S/N$ of 3.1). Unfortunately, the narrowband images were
too insensitive to show such faint stellar sources, which therefore
ruled out confirmation of the detection of this point-like source in
another band. In Fig.~\ref{hst_radio_newIDs}, where we plot the F547M
image superimposed with the detected point-like sources that survived
our DAOPHOT-II/allstar analysis, it is clear that object 1$_{HST}$
lies within the compact nebulosity which is unambiguously associated
with the core of the CTB 80 SNR. It is also clear that this object is
within the combined radio/optical error region for PSR B1951+32.

However, there is a problem with this photometric result for object
1$_{HST}$. An inspection of the resulting star-subtracted image shows
that MAG$_{F547M}$ = 24.3 must be an overestimation of the flux of the
apparent point source within the nebula, as the
fitting-and-subtraction process does not result in a smooth underlying
background, but rather a pronounced ``hole'' in the nebula. This is a
consequence of three probable factors: the undersampling of the WF3
PSF, coupled with the rapidly changing background in the vicinity of
the source, and the possibility that the source is not a pure
point-source but may be broadened by being embedded in a knot of the
nebulosity.

\begin{figure}
\resizebox{\hsize}{!}{\includegraphics{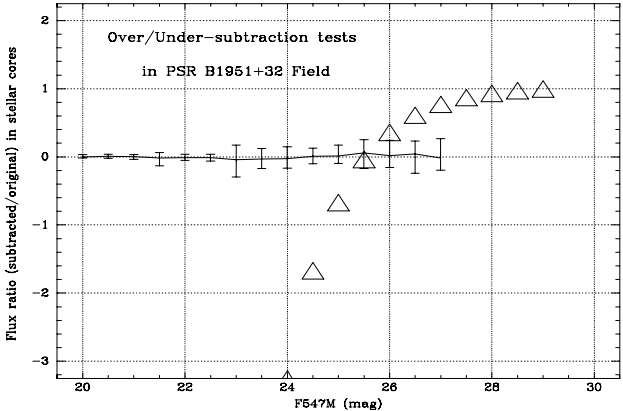}}

\caption{Results of the star-subtraction tests to determine the
magnitude range of the pulsar counterpart candidate 1$_{HST}$ wherein
quantifiable over-subtraction or under-subtraction does not occur.
Positive values of the flux ratio indicate under-subtraction (with +1
being essentially no subtraction at all), and negative values indicate
over-subtraction. The triangular symbols denote the results for the
pulsar counterpart candidate's test magnitudes. The line trace and
error bars denote the median values and 1-$\sigma$ errors of all the
well-measured stars (those whose aperture photometry and PSF-fitting
photometry agree to within $\pm$0.5 mag) over the entire WF3 chip, binned
in 0.5-mag intervals.}
\label{oversubtract}
\end{figure}

\begin{figure}
\resizebox{\hsize}{!}{\includegraphics{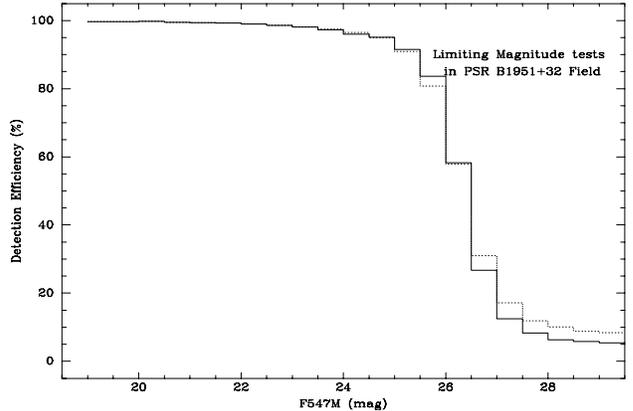}}
\caption{Results of the Monte Carlo simulations for star
detection/measurement probabilities in the 512x512 pixel region
centred on the nominal pulsar position. The heavy, dotted trace is for
the original F547M image and the lighter trace is for the case of
increased background. The background ``detection'' level seen here, at
much fainter magnitudes than the realistic limiting magnitude cutoff,
is due to the approximately 8\% probability of ``detecting'' either noise peaks, or an {\it
existing} star which happens to lie within the 2-pixel
matching tolerance of an artificial star.}
\label{mag_limits}
\end{figure}

To investigate this issue we performed the following test. The
position of the source 1$_{HST}$, as determined by DAOPHOT-II/allstar,
was kept constant, while its magnitude was varied from MAG$_{F547M}$ =
21 - 29 in steps of 0.5 magnitude. At each magnitude it was subtracted
from the original image and aperture photometry was performed on the
resulting subtracted image, in order to measure the residual flux
level within the central core pixels of the source. These residual
fluxes were plotted as a ratio of the flux measured using the same
aperture and background annulus on the original image; this ratio is
an estimate of the subtraction smoothness - see
Fig.~\ref{oversubtract}. In this Figure, we also plot the median
values and error bars for the same flux ratio of all the well-measured
stars over the entire WF3 chip, binned in 0.5-mag intervals. For our
pulsar counterpart candidate, one would expect to see subtraction
smoothness of the same order as had been achieved for stars of the
same brightness elsewhere in the field. The conclusion is that, to
first order, the source would have to have MAG$_{F547M}$ $\ge$ 25.2 to
prevent {\it oversubtraction} into the nebula; and {\it
undersubtraction} would occur at MAG$_{F547M}$ $\ge$ 26.2. The
magnitude of object 1$_{HST}$ therefore lies within this interval, but
its $S/N$ would be reduced to $\approx$2 - 2.5.

The question then arises, could one normally detect such a faint
source as object 1$_{HST}$ in this data without these corrections, or
is it spurious? We performed Monte Carlo simulations in order to test
this hypothesis. Using the PSF shape and photometric zeropoint
previously determined for the F547M image, artificial stars were added
into the 512x512 pixel region centred on the nominal pulsar
position. For each 0.5 magnitude step from MAG$_{F547M}$ = 21 - 29, 25
stars were added to this field at random positions, and then the same
automatic star detection and measurement process was repeated in order
to try to recover them. This entire process was repeated 100 times for
robust statistics on the efficiency of star detection with increasing
magnitude. Then, to take into account the effect of the higher
background within the small nebula, which increases the sky noise, the
simulations were repeated with the background everywhere increased to
the mean background within the small nebula. The results are shown by
the two traces in Fig.~\ref{mag_limits}. Clearly, there is a good
(between $\approx$ 60\% - 80\%) probability of detecting a point source as
faint as MAG$_{F547M}$ = 25.5 - 26.5. The {\it level} of the increased
nebular background is shown to have only a very marginal effect on
star detection (although the {\it gradients} within the nebular
background are likely causes for bias in the photometric measurement,
as has already been noted).

Crucially, the estimated magnitude of MAG$_{F547M}$ = 25.2 - 26.2 for
object 1$_{HST}$, including the correction for the oversubtraction
effect, is within the range predicted by the successful mode framework
of Pacini \& Salvati (1987) and more recently the phenomenological
analysis of Shearer \& Golden (2001).  The magnitude of object
4$_{HST}$ is also consistent with these models, at the brighter end of
the predicted range. Consequently, taken together with their positions
with respect to the two radio-position error ellipses, we suggest that
these two objects are plausible new optical counterpart candidates to
PSR B1951+32. Both candidates lie within a region of area $\approx$ 1
arcsec$^{2}$. To estimate the probability of unrelated field stars
being found within this region and at the same magnitude as the
candidates, we computed a luminosity function (LF) for the total
measured stellar content of the three near-identical WF chips - WF3
containing the pulsar field, and WF2 and WF4 abutting it and
containing similar starfields. The LF was then normalised to give the
density of foreground/background objects per arcsec$^{2}$ at 0.5-mag
intervals. For object 4$_{HST}$ at MAG$_{F547M}$ = 24.5, the LF has
over 300 stars or 0.02 stars per arcsec$^{2}$.  For object 1$_{HST}$
at MAG$_{F547M}$ $\approx$ 25.75, the LF has over 700 stars or 0.05
stars per arcsec$^{2}$. These numbers were not significantly increased
when the LF was corrected for the corresponding detection efficiencies
shown in Fig.~\ref{mag_limits}. For the entire magnitude range of
MAG$_{F547M}$ = 24 - 26.5, generously bracketing the two candidates and their
uncertainties, the LF has over 3000 stars or 0.18 stars per
arcsec$^{2}$.  Thus the probability of finding two such objects an
arcsecond apart is significant, but low.

%______________________________________________________________

\section{Conclusions}

\begin{figure}
\resizebox{\hsize}{!}{\includegraphics{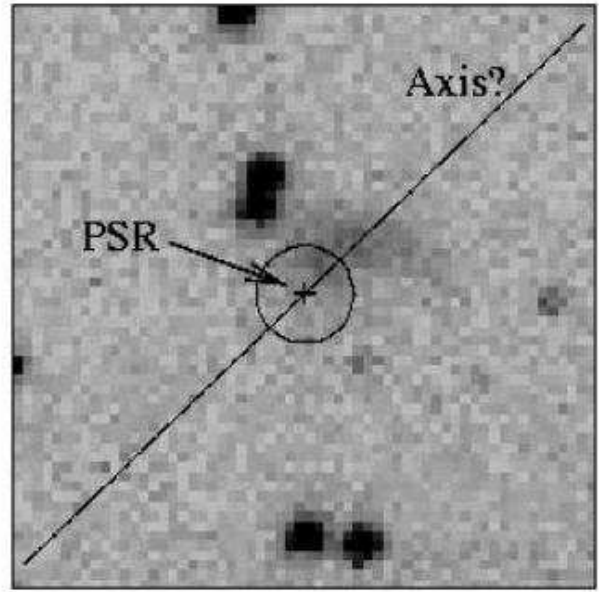}}
\caption{Location of the radio counterpart according to Hester (2000b),
along an axis that bisects apparent Balmer-dominated lobes (which lie
outside the field shown here) in relation to the synchrotron
nebula. Figure adopted from Hester (2000b).}
\label{hst_hester}
\end{figure}

Hester (2000a) has reported on a similar analysis of the same archival
HST data, and agrees that the knot of extended continuum emission is
synchrotron dominated as a consequence of the pulsar wind.  Hester
(2000b) places the radio counterpart 0.5'' to the East/SE of this
nebula (see Fig.~\ref{hst_hester}) - roughly between our proposed
counterpart candidates 1$_{HST}$ and 4$_{HST}$, although closer to
1$_{HST}$ than to 4$_{HST}$. We have reproduced his results by mapping
the Foster et al. (1994) radio timing position onto the {\it original}
GSC-based astrometric solution from the HST data pipeline.  However,
we are more confident in our 2MASS-recalibrated and astrometry, which
also benefits from our correction utilising the accurate proper motion
rate, published some time after Hester's analysis by Migliazzo et
al. (2002). As a result our mapped radio timing position differs from
Hester's by 0$\farcs$47. Hester also comments that the relative
locations, orientation and luminosity are similar to the knot situated
0.5'' to the SE of the Crab pulsar. The agreement, in terms of
synchrotron luminosity from the nebula, is compelling, regardless of
which of the two astrometric solutions are used, since they overlap
within their errors. However the location of the pulsar is still
uncertain, given the considerable discrepancy between the two best
independent radio positions.  We have outlined the evolution in these
radio positions and optical astrometric solutions which, taken
together, locate the pulsar in an error region which is still too
large to definitively pin down the pulsar's optical counterpart;
however the more accurate mapped radio position (in terms of formal
quoted errors) lies to the East/SE of the nebula. This supports the
geometric orientation suggested by Hester (2000a,b).

One of the two newly resolved sources which we have identified as
1$_{HST}$ and 4$_{HST}$ may therefore indeed be the $\sim$ 24th-26th
magnitude pulsar as predicted by the models of Pacini \& Salvati
(1987) and Shearer \& Golden (2001). The other possibilities are that
both are unrelated field stars, or that 1$_{HST}$ is a localised
`knot' within the bigger `knot' of the synchrotron nebula while
4$_{HST}$ is a field star; both of these possibilities would imply
that the true optical counterpart is fainter still. This would
possibly reconcile one difficulty with the present analysis - the
inconsistency of the two new candidates with the formally small error
ellipse of the radio timing position; but it would create another
difficulty - the inconsistency of candidates fainter than 26th
magnitude with the phenomenological model predictions. Deeper HST
and/or diffraction-limited adaptive optics imaging will be needed to
determine which of these various possibilities is the correct
interpretation, by putting some hard constraints on the colour and
spectral index of both the synchrotron nebulosity and the candidate
counterparts. Goldoni et al. (1995) noted that the optical spectral
index {\it steepened} with age for the known optically emitting
pulsars, while Shearer \& Golden (2001) showed that this is consistent
with the flattening of the pulse-peak luminosity relationship with the
outer field strength. Consequently, with its expected steep optical
spectral index, one would expect that the optical counterpart to PSR
B1951+32 will not be as distinctly blue as younger Crab-like pulsars,
but nevertheless distinguishable from regular thermally-emitting
stars. Additionally, one could test the candidates by looking for
proper motion, which (since the epoch of these HST observations)
should now amount to a possibly detectable
$0^{\prime\prime}.12$. Follow-up timing studies with the upgraded
TRIFFID imaging photon-counting camera would permit the search for
optical pulsations from a securely identified counterpart, with much
higher statistical significance than previously. Not only would this
provide the essential confirmation of any optical counterpart, it
would also test the hitherto successful optical models of Pacini \&
Salvati (1987) and Shearer \& Golden (2001) in an age regime where few
optical pulsars have been found.

%______________________________________________________________

\begin{acknowledgements}

The authors gratefully acknowledge financial support from Enterprise
Ireland under the Basic Research Programme. RFB is also grateful for
financial support from the Improving Human Potential programme of the
European Commission (contract HPFM-CT-2000-00652). This publication is
based upon Hubble Space Telescope data obtained from the ST-ECF
archive, ESO, Garching, Germany. It also makes use of data products
from the Two Micron All Sky Survey, which is a joint project of the
University of Massachusetts and the Infrared Processing and Analysis
Center/California Institute of Technology, funded by the National
Aeronautics and Space Administration and the National Science
Foundation. Finally, we also wish to thank the referee, Richard Strom,
for his useful suggestions and constructive comments which have
considerably improved the paper's content and presentation.

\end{acknowledgements}

%______________________________________________________________

%______________________________________________________________

\end{document}